%% file: main.tex
\documentclass[sigconf]{acmart}

\makeatletter
\def\@ACM@checkaffil{
    \if@ACM@instpresent\else
    \ClassWarningNoLine{\@classname}{No institution present for an affiliation}%
    \fi
    \if@ACM@citypresent\else
    \ClassWarningNoLine{\@classname}{No city present for an affiliation}%
    \fi
    \if@ACM@countrypresent\else
        \ClassWarningNoLine{\@classname}{No country present for an affiliation}%
    \fi
}
\makeatother

\AtBeginDocument{%
  \providecommand\BibTeX{{%
    \normalfont B\kern-0.5em{\scshape i\kern-0.25em b}\kern-0.8em\TeX}}}

\copyrightyear{2023}
\acmYear{2023}
\setcopyright{rightsretained}
\acmConference[CIKM '23]{Proceedings of the 32nd ACM International Conference on Information and Knowledge Management}{October 21--25, 2023}{Birmingham, United Kingdom}
\acmBooktitle{Proceedings of the 32nd ACM International Conference on Information and Knowledge Management (CIKM '23), October 21--25, 2023, Birmingham, United Kingdom}
\acmDOI{10.1145/3583780.3615061} 
\acmISBN{979-8-4007-0124-5/23/10}

\usepackage{enumitem}
\usepackage{stmaryrd}
\usepackage{subfig}
\usepackage{amsmath,enumitem}
\usepackage[ruled,vlined,linesnumbered,noresetcount]{algorithm2e}
\usepackage{threeparttable,color}
\usepackage{booktabs}
\usepackage{multirow}
\usepackage{multicol}
\usepackage{soul}

\newcommand{\method}{\textit{scMoFormer}\xspace}

\begin{document}

\title{Single-Cell Multimodal Prediction via Transformers}

\author{Wenzhuo Tang}
\authornote{Equal contribution.}
\email{tangwen2@msu.edu}
\orcid{0000-0002-7038-5765}
\affiliation{%
  \institution{Michigan State University}
}

\author{Hongzhi Wen}
\email{wenhongz@msu.edu}
\orcid{0000-0003-0775-8538}
\authornotemark[1]
\affiliation{%
  \institution{Michigan State University}
}

\author{Renming Liu}
\email{liurenmi@msu.edu}
\orcid{0000-0002-6025-6492}
\authornotemark[1]
\affiliation{%
  \institution{Michigan State University}
}

\author{Jiayuan Ding}
\email{dingjia5@msu.edu}
\orcid{0000-0001-5783-0062}
\affiliation{%
  \institution{Michigan State University}
}

\author{Wei Jin}
\email{wei.jin@emory.edu}
\orcid{0000-0002-5054-954X}
\affiliation{%
  \institution{Emory University}
}

\author{Yuying Xie}
\email{xyy@msu.edu}
\orcid{0000-0002-1049-2219}
\affiliation{%
  \institution{Michigan State University}
}

\author{Hui Liu}
\email{liuhui7@msu.edu}
\orcid{0000-0002-3555-3495}
\affiliation{%
  \institution{Michigan State University}
}

\author{Jiliang Tang}
\email{tangjili@msu.edu}
\orcid{0000-0001-7125-3898}
\affiliation{%
  \institution{Michigan State University}
}

\renewcommand{\shortauthors}{Tang et al.}

\begin{abstract}
    \input{sections/abstract.tex}

\end{abstract}



\maketitle

\input{sections/intro.tex}

\input{sections/related_works.tex}

\input{sections/problem_setup.tex}

\input{sections/method.tex}

\input{sections/experiment.tex}

\input{sections/conclusion.tex}

\input{sections/acknowledgments}

\appendix
\input{sections/appendix.tex}
\bibliographystyle{ACM-Reference-Format}
\bibliography{main}


\end{document}

%% file: sections/abstract.tex
The recent development of multimodal single-cell technology has made the possibility of acquiring multiple omics data from individual cells, thereby enabling a deeper understanding of cellular states and dynamics. 
Nevertheless, the proliferation of multimodal single-cell data also introduces tremendous challenges in modeling the complex interactions among different modalities. The recently advanced methods focus on constructing static interaction graphs and applying graph neural networks (GNNs) to learn from multimodal data. However, such static graphs can be suboptimal as they do not take advantage of the downstream task information; meanwhile GNNs also have some inherent limitations when deeply stacking GNN layers. To tackle these issues, in this work, we investigate how to leverage transformers for multimodal single-cell data in an end-to-end manner while exploiting downstream task information. In particular, we propose a \method{} framework which can readily incorporate external domain knowledge and model the interactions within each modality and cross modalities.
Extensive experiments demonstrate that \method{} achieves superior performance on various benchmark datasets. Remarkably, \method{} won a Kaggle silver medal with the rank of $24 / 1221$ (Top 2\%) {\it without ensemble} in a NeurIPS 2022 competition\footnote{\url{https://nips.cc/virtual/2022/competition/50092}}. Our implementation is publicly available at Github\footnote{\url{https://github.com/OmicsML/scMoFormer}}.


%% file: sections/intro.tex
\section{Introduction}

Advancements in multimodal single-cell technologies provide the capability to simultaneously profile multiple data types in the same cell, including chromatin accessibility~\cite{cao2018joint, chen2019high}, DNA methylation~\cite{gaiti2019epigenetic}, nucleosome occupancy~\cite{pott2017simultaneous}. These technologies offer an exciting opportunity to characterize cell identity and state at an unprecedented resolution, enabling a better understanding of gene regulatory networks in multicellular organisms and tissues~\cite{zhu2020single}.  Despite the rapid accumulation of multimodal single-cell data, the analyses of such data are still faced with numerous challenges. First, single-cell measurements often exhibit high sparsity and noise levels, making it difficult to draw meaningful insights from the data~\cite{eraslan2019single}. 
Furthermore, samples are often measured under different conditions, including batches, times, locations, or using different instruments, which leads to systematic variations in the measured values, which further complicates the interpretation of single-cell data. These imperfections can result in biased estimates of cell-cell interactions and poses tremendous challenges for computational models to exploit such interactions.

To effectively capture the intricate interactions within cells and genes, current research focuses on constructing static graphs based on heuristic criteria and then employing graph neural networks (GNNs)~\cite{kipf2016semi, velickovic2017graph, ma2021deep} to extract information from the built graph. 
However, the quality of built graphs is contingent upon the selected heuristic similarity measure and it does not incorporate downstream task information.
An alternative way for building the interaction graph is to construct the graph based on domain knowledge, e.g., utilizing publicly accessible databases~\cite{wen2022graph}. 
However, similar to k-NN graphs, such graph construction process does not leverage downstream task information and the knowledge base may not include all relevant genes/proteins. Furthermore, GNNs have some inherent limitations that hinder their success in applications: the over-smoothing~\cite{kreuzer2021rethinking} and over-squashing~\cite{alon2021on} issues where GNNs produce poor results when we deeply stack GNN layers.
Hence, one question naturally arises: \textit{can we have a better approach to construct interaction graphs (among cells, genes, and proteins) that utilize downstream information while avoiding the aforementioned issues?}

In light of the recent advances of transformers~\cite{devlin2018bert, kitaev2020reformer, liu2021swin, liu2022swin} in capturing pairwise relations among objects, we seek to utilize transformers for learning the interaction graph for cells, genes, and proteins in an end-to-end manner. Transformers are well-suited to address the limitations of static graphs: they learn the interaction between objects through the self-attention mechanism, where all objects are attended to each other with learnable attention scores indicating their interaction strength. Thus, the attention matrix provides an advanced approach to characterize the interaction between objects in a data-driven way and has demonstrated success in reducing unwanted variance and noise across batches~\cite{yang2022scbert}. 
However, these traditional transformers do not account for the available graph structure and are therefore unable to leverage prior information present in graph data, such as biological knowledge graphs. In this context, graph transformers~\cite{rampasek2022recipe, ying2021transformers, dwivedi2020generalization}  offer a solution by combining the strengths of GNNs and transformers to make use of graph data. These approaches allow for the incorporation of prior insights from structural information learned from GNNs, while still allowing for data-specific interactions to be learned through the attention mechanism.
On top of that, graph transformers also alleviate the over-smoothing and over-squashing problems in GNNs by enabling individual objects to attend to unconnected objects~\cite{chen2022structure, dwivedi2021graph, kreuzer2021rethinking, rampavsek2022recipe}. 
Therefore, it is of great importance to investigate the potential of (graph) transformers in single-cell analysis.

In this work, we aim to design a transformer framework for multimodal single-cell  data. In essence, to utilize the strengths of (graph) transformers, we are faced with two non-trivial challenges.  \textbf{First}, since multimodal data contains diverse information from various sources, e.g., genes, proteins, and cells, it can be difficult for a single transformer to capture all aspects. \textbf{Second}, traditional transformers suffer a quadratic computation complexity w.r.t. the number of objects, which poses a challenge for single-cell analysis where the number of cells can be large. 
To address the first challenge, we introduce the \underline{S}ingle-\underline{C}ell Multi\underline{mo}dal Trans\underline{former} \method{}, which employs multiple transformers to model the multimodal data, allowing each transformer to deal with a specific data modality. The core of \method{} is the cross-modality aggregation component which builds a bridge between these transformers and aggregates the necessary information from individual ones.
For the second challenge, \method employs linearized transformers~\cite{choromanski2021rethinking} to the cells which greatly reduces the computational complexity.
To the best of our knowledge, we are the first to employ transformers to advance the analysis of multimodal single-cell  data. Our proposed framework achieves promising results on the benchmark datasets, providing a very strong baseline for follow-up research. Our contributions can be summarized as follows: 
\vspace{-1em}
\begin{itemize}[leftmargin=*]
    \item We study the problem of multimodal single-cell data analysis and propose a transformer framework \method{} to capture the intricate relations within modalities and between modalities. 
    \item The proposed \method{} is versatile and can flexibly incorporate domain knowledge regarding genes and proteins. 
    \item The proposed \method{} achieves superior performance on various benchmark datasets. Remarkably, we are one of the top winners in a NeurIPS 2022 competition. 
\end{itemize}


%% file: sections/related_works.tex
\section{Related Work}
In this section, we review related works of our proposed framework, including GNNs, transformers, and other in single-cell analysis.

\subsection{Deep Learning on Multimodal Integration} 
There is a growing number of deep learning-based methods for multimodal single-cell analysis in the community. For instance,
scMDC~\cite{lin2022clustering} is an end-to-end autoencoder-based model with one encoder and two decoders. The encoder takes the concatenation of two modalities as an input and then reconstructs two modalities separately via two individual decoders.
DCCA~\cite{zuo2021deep} learns a coordinated but distinct representation for each omics data by mutually supervising each other on the basis of semantic similarity across embeddings, and then reconstructs back to the original dimension as output via a decoder for each omics data.
Cross-modal Autoencoders~\cite{yang2021multi} utilize multiple autoencoders to map different modalities onto the same latent space, and incorporate prior knowledge through the use of adversarial loss and paired anchor loss in the training process. 
BABEL~\cite{wu2021babel} consists of two neural-network-based encoders and two decoders for translation between gene expression and chromatin accessibility. Both Cross-modal Autoencoders and BABEL focus on multimodal translation by adding interoperability constraints to train multiple encoders and decoders. 
Another approach, scMM~\cite{minoura2021mixture}, captures nonlinear latent structures with variational autoencoders. It exploits a mixture-of-expert framework with a deep generative model and attains end-to-end learning by modeling raw counts of each modality. While these models have made significant advancements in multimodal integration, most of them are based on autoencoders and tend to overlook the underlying biological interactions of molecules and cells.

\subsection{GNNs and Transformers in Single-Cell} 
To capture the biological interactions of molecules and cells, there has been an increasing number of GNNs and transformer frameworks published in the field of single-cell analysis. 
One benefit of transformers applied in single-cell data is to capture long-range dependency in a global view. Another benefit is to interpret biological phenomena via the attention mechanism in transformers.
From GNNs' perspective, graphs are natural to represent all kinds of data in single-cell data, like gene-to-gene graphs, cell-to-cell graphs, and cell-to-gene graphs. Another benefit of GNNs is to easily add prior knowledge into graphs, like pathways between genes or overlaps between genes and peaks. 
For example, scGNN~\cite{wang2021scgnn} models cell-cell interaction by incorporating GNN with multi-modal autoencoders. Specifically, scGNN builds a cell graph by capturing cell-type-specific regulatory signals and utilizes a Left-Truncated Mixture Gaussian model for scRNA-Seq data. GLUE~\cite{cao2022multi} pre-trains modality-specific variational autoencoders to get cell embeddings and then encodes a knowledge-based graph with GNNs. The next step involves performing an adversarial multimodal alignment of the cells through an iterative optimization process. In addition, ScMoGNN~\cite{wen2022graph} models the cell similarity and feature similarity by building a cell-feature graph and extracts information from data with a graph encoder. ScMoGNN takes advantage of gene pathway data as prior knowledge to enhance the graph and denoise the data. Moreover, scBERT~\cite{yang2022scbert} follows the pre-training and fine-tuning paradigm of bidirectional encoder representations from transformers (BERT) for cell annotation of scRNA-seq data. The process of annotation involves extracting high-level patterns of cell types from the reference dataset. Different from these approaches which focus on single-modality data,  we are the first to introduce transformers and GNNs to single-cell multimodal prediction. 

%% file: sections/problem_setup.tex
\section{Problem Statement}

Before we state the problem, we first introduce the notations used in the following sections. For clarity and simplicity, we use the subscripts "g", "p" and "c" for gene, protein, and cell, respectively. For instance, we use $\mathbf{h}_g$, $\mathbf{h}_p$, $\mathbf{h}_c$ to denote the embeddings of genes, proteins and cells, respectively.

In this work, we follow the problem setting in the NeurIPS 2022 competition, i.e., Multimodal Single-Cell Integration Across Time, Individuals, and Batches\footnote{https://www.kaggle.com/competitions/open-problems-multimodal/}. The goal of this competition is to predict a paired modality with a given modality and to infer how DNA, RNA, and protein measurements co-vary in single cells. Specifically, we focus on using gene expression (RNA) to predict surface protein level. We denote $\mathbf{X}_{\text{g}}$ and $\mathbf{X}_{\text{p}}$ as the measurement counts of gene and protein, respectively. With $\mathbf{X}_{\text{g}}$, we try to learn a mapping function that can best describe the relationship between two modalities. We denote $\mathcal{L} \left(\cdot,\cdot\right)$ as the objective function that measures the dissimilarity between the predicted and the true protein level. Formally, we describe our target as an optimization problem:

\textit{Given $\mathbf{X}_{\text{g}}$ and the objective function $\mathcal{L}\left(\cdot,\cdot\right)$, we aim to find a mapping function $f_{\theta}^*$ (parameterized by $\theta$) that minimize the objective loss:
\begin{equation}
    f_{\theta}^* = \arg\min_{f_{\theta}} \mathcal{L} \left( f_{\theta} \left(\mathbf{X}_{\text{g}}\right), \mathbf{X}_{\text{p}}\right).
\end{equation}}

In the subsequent sections, we formulate the mapping $f_{\theta}^*$ using transformers and GNNs and employ Root Mean Square Error, Mean Absolute Error, and Pearson correlation coefficient as evaluation metrics for our predictions.

%% file: sections/method.tex
\section{Our Multimodal transformer Framework}

\begin{figure*}[t]
    \centering
    \vspace{-1em}
    \includegraphics[width=0.8\linewidth]{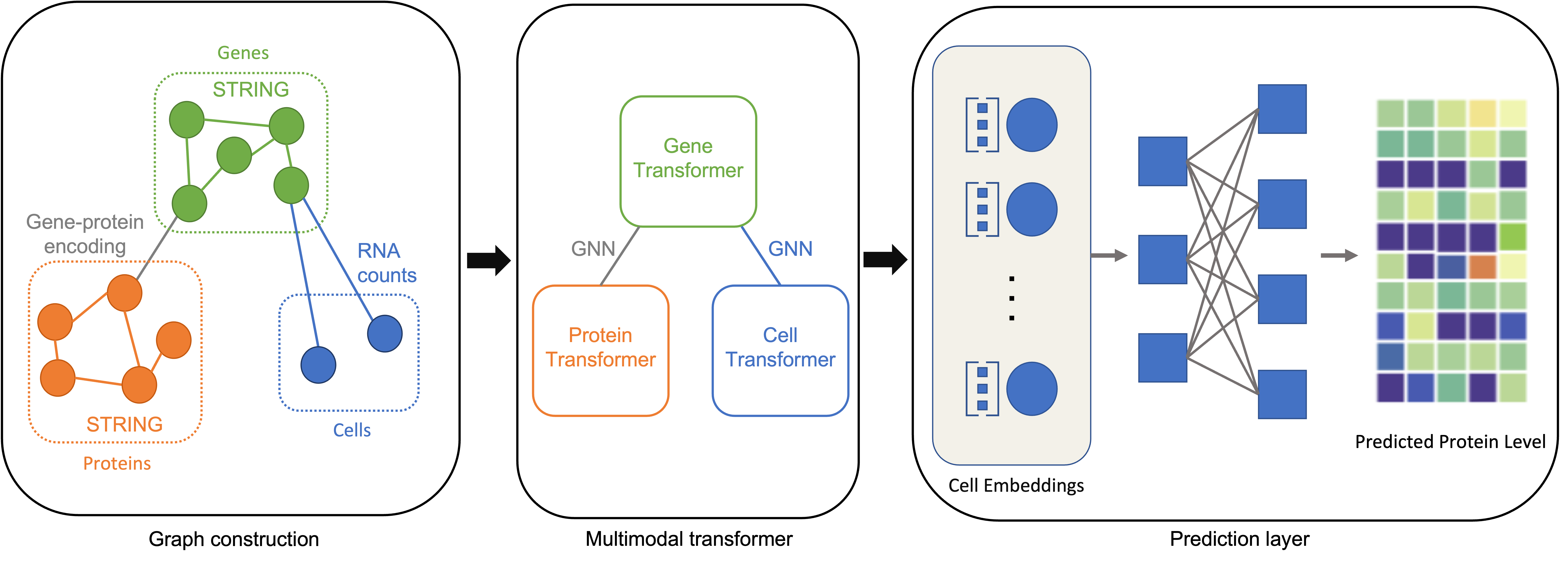}
    \vspace{-1.5em}
    \caption{An illustration of \method{}. In this framework, three important components are included: graph construction, multimodal transformer, and prediction layer.}
    \label{fig:framework}
    \vspace{-1.5em}
\end{figure*}

We now introduce the proposed multimodal framework \method{}. An shown in Figure~\ref{fig:framework}, it consists of multimodal graph construction, a multimodal transformer, and a prediction layer. In brief, we first construct a heterogeneous graph that contains cell, gene, and protein nodes together with their interactions. We then utilize multiple (graph) transformers to extract rich cell representations and predict each cell's surface protein abundance levels. 

\vspace{-0.4em}
\subsection{Multimodal Graph Construction}
\label{sec:graph_const}



In this subsection, we introduce how we construct the graph for multimodal single-cell data. Specifically, we construct a heterogeneous graph containing three different types of nodes. It has four subgraphs as shown in Figure~\ref{fig:framework}, i.e., a protein-protein graph, a gene-gene graph, a gene-protein graph, and a cell-gene graph. Next, we detail how to conduct these subgraphs. 

\subsubsection{Subgraph Construction}\label{sec:subgraph} 

\noindent \textbf{Protein-Protein Graph.} 
To integrate prior biological information into protein-protein graph, we refer to the STRING~\cite{szklarczyk2023string} database. STRING provides a comprehensive resource of protein-protein interactions and the functional relationships between different proteins. It contains seven different channels covering varied aspects of sources, including genomic context, experiment results, and text mining efforts. We use the combined confidence score of all channels as edge weights to comprehensively enhance the graph. The proteins in STRING are labeled in Ensemble~\cite{cunningham2022ensembl} protein (ENSP), and the mapping from ENSP to the protein preferred name is provided in additional information resource. Notice that regularly one protein may have multiple aliases. To match the protein display names with ENSP, we utilize GeneCards~\cite{stelzer2016genecards} to find all possible aliases of our target proteins. In a nutshell, the mapping from STRING nodes to our target proteins is given by ENSP $\shortrightarrow$ possible aliases of proteins $\shortrightarrow$ the display names of proteins within the dataset.

\noindent\textbf{Gene-Gene Graph.} To maintain consistency and eliminate potential noise from multiple sources of prior information, we also utilized the STRING database to construct the gene-gene graph and combined all seven channels of the STRING database to enhance the graph as much as possible. Note that although the gene and protein nodes are biologically equivalent in the sense of prior information, we process them separately to more specifically handle data-specific information related to the target labels. The genes are labeled in Ensemble gene (ENSG) and additional matching efforts are needed to form the gene-gene graph. We utilize the MyGeneInfo~\cite{wu2014mygene} gene query service to align the STRING protein nodes, encoded by ENSP, with the input gene nodes, encoded by ENSG.

\noindent\textbf{Gene-Protein Graph.} Now we have two separate graphs among proteins and genes respectively, and we would like to form a general frame by adding the connections between genes and proteins. Following the central dogma of molecular biology, information flows from RNA to proteins via translation. This encoding relationship is recorded within the gene and protein nomenclature. Specifically, the gene names from existing single-cell multimodal datasets contain two parts: the ENSG and the symbol or name of corresponding proteins. That is, if a gene and a protein share the same symbol, it means the target protein is encoded by the gene. Utilizing biological information, we link the proteins to genes by matching their symbols.

\noindent\textbf{Cell-Gene Graph.} The constructed gene-protein graph is fully based on prior knowledge, and we integrate data-based information into the multimodal heterogeneous graph by involving cell nodes. The gene expression counts of multimodality data imply the data-specific relationships between genes and cells. Naturally, a cell and a gene are connected if the gene expresses within that cell. Note that the number of genes detected within each cell varies significantly, which depicts that the raw counts of RNA abundance show substantial heterogeneity among cells. In addition, the raw data includes extremely large counts, making it impractical to apply counts data as the edge weights between cells and genes. Thus, we normalize each cell by total counts over all genes and take the logarithm. 
The processed data is then fed into the multimodal graph to describe the cell-gene links.

\noindent\textbf{Remark.} 
It is worth noting that we do not construct the cell-cell graph. Instead, we use the transformer to learn the cell-cell relationships via the attention mechanism. This is in contrast with some previous studies that utilize a static cell-cell similarity graph generated from the input features, which might be prone to inaccurate cell relationships due to the noisy nature of single-cell data. On the other hand, we do not explicitly establish connections between the cells and the target protein nodes, as it might cause the model to easily overfit. More discussions are included in Appendix~\ref{app:remark}.
\subsubsection{Heterogeneous Graph Construction} 

\noindent\textbf{Formal Graph Definition.} We now formally define the multimodal heterogeneous graph. Let $\mathbf{A}$ be the adjacency matrix of the multimodal heterogeneous graph with $\mathcal{V} = \left(\mathbf{v}_{\text{p}}, \mathbf{v}_{\text{g}}, \mathbf{v}_{\text{c}} \right)$ as the node set, 
and $N_{\text{p}}$, $N_{\text{g}}$, $N_{\text{c}}$ are the number of proteins, genes, cells, respectively. Denote $\mathcal{E}\in \mathcal{V} \times \mathcal{V}$ as the edge set, we write the multimodal heterogeneous graph as $\mathcal{G} = \left(\mathcal{V}, \mathcal{E} \right)$. Let $N =(N_{\text{p}}+N_{\text{g}}+N_{\text{c}})$ be the total number of nodes, the adjacency matrix $\mathbf{A}$ can be written as follows:
\begin{equation}
    \mathbf{A}=\left(\begin{array}{ccc}
        \mathbf{A}_{\text{p}} & \mathbf{S}^T & \mathbf{0} \\
        \mathbf{S} & \mathbf{A}_{\text{g}} & \mathbf{A}_\text{RNA}^T \\
        \mathbf{0} & \mathbf{A}_\text{RNA} & \mathbf{0} \\
    \end{array}\right) \in \mathbb{R}^{N\times N},
\end{equation}
where $\mathbf{A}_{\text{p}} \in \mathbb{R}^{N_{\text{p}}\times N_{\text{p}}}$ is the protein-protein interaction graph; $\mathbf{A}_{\text{g}} \in \mathbb{R}^{N_{\text{g}}\times N_{\text{g}}}$ denotes the gene-gene graphs mapped from STRING via MyGene.info~\cite{lelong2022biothings}; $\mathbf{S} \in \mathbb{R}^{N_{\text{g}}\times N_{\text{p}}}$ is the encoding relationship between genes and proteins; and $\mathbf{A}_\text{RNA} \in \mathbb{R}^{N_{\text{c}}\times N_{\text{g}}}$ represents the gene expression. 

\noindent\textbf{Node Feature Initialization.} With the graph structure, next we discuss how to initialize the node features. 
Since the RNA counts $\mathbf{X}_\text{g}$ show drastic sparsity along with high dimension, it is not practical to be directly used as cell features. Hence, we first denoise the data and reduce the dimension by Singular Value Decomposition (SVD). To alleviate the effects of cell-to-cell heterogeneity and extremely large counts, we conduct library size normalization and centered log-transformation. The preprocessed data $\bar{\mathbf{X}}$ is then passed through the SVD algorithm. Cell features are initialized with the reduced features $\mathbf{h}^0_{\text{c}} \in \mathbb{R}^{N_{\text{c}}\times d_0}$. We then initialize gene features by the weighted sum of the reduced features $\mathbf{h}^0_{\text{g}} = \bar{\mathbf{X}}_\text{g}^T \cdot \mathbf{h}^0_{\text{c}} \in \mathbb{R}^{N_{\text{g}}\times d_0}$ with the normalized counts $\bar{\mathbf{X}}_\text{g}$ as the weights. In the studied problem, proteins are the target modality for prediction; thus they are initialized randomly based on their indices.

\subsection{Multimodal Transformers}
To effectively handle the heterogeneity in the heterogeneous graph, we will introduce how \method{}  employs multiple transformers to process each individual modality and utilizes a cross-modality aggregation mechanism to combine the information.


\subsubsection{Cell Transformer}
Transformer~\cite{vaswani2017attention} has made significant achievements in the field of Natural Language Processing (NLP) in recent years. The attention mechanism can capture high-order and non-Euclidean connections between nodes, which is desired to explore the cell-cell relationships within single-cell multimodal data. We denote the queries, keys, and input cell embeddings as $\mathbf{Q}$, $\mathbf{K}$, $\mathbf{h}_{\text{c}} \in \mathbb{R}^{N_{\text{c}}\times d}$ with input dimension $d$, 
the scaled dot-product attention can be formulated as $\operatorname{Attn}(\mathbf{h}_\text{c})=\operatorname{softmax}\left({\mathbf{Q} \mathbf{K}^{T}}/{\sqrt{d}}\right) \mathbf{h}_\text{c}$,
where $\mathbf{A}_{\text{attn}} = \operatorname{softmax}\left({\mathbf{Q} \mathbf{K}^{T}}/{\sqrt{d}}\right)$ is the attention matrix of cell-cell interactions. Despite being effective in various NLP tasks, the original transformer has limitations in scalability due to its relatively high space complexity of $O(N_\text{c}^2 + N_\text{c} d)$ and time complexity of $O(N_\text{c}^2·d)$. This issue considerably limits the application of transformers in single-cell multimodal analysis. Since typically multimodal data includes tens of thousands of cells, it is not applicable to implement the original transformer directly on cells. 

To address the scalability issue, we employ generalized kernelizable attention~\cite{choromanski2021rethinking} as a computationally efficient approximation of traditional attention. The attention blocks is kernelized in the form $\mathbf{A}(i, j)=\mathrm{K}\left(\mathbf{q}_{i}^{\top}, \mathbf{k}_{j}^{\top}\right)$, where $\mathbf{q}_{i}$ stands for the $i^{t h}$ row of query $\mathbf{Q}$ and $\mathbf{k}_{j}$ denotes the $j^{t h}$ row of key $\mathbf{K}$. The kernel $\mathrm{K}: \mathbb{R}^{d} \times \mathbb{R}^{d} \rightarrow \mathbb{R}_{+}$ is specified as $\mathrm{K}(\mathbf{x}, \mathbf{y})=\mathbb{E}\left[\phi(\mathbf{x})^{\top} \phi(\mathbf{y})\right]$,
where $\phi: \mathbb{R}^{d} \rightarrow \mathbb{R}_{+}^{r}$ denotes the feature mapping. Let $\mathbf{Q}^{\prime}$, $\mathbf{K}^{\prime} \in \mathbb{R}^{N_\text{c} \times r}$ be the approximate query and key with rows given as $\phi\left(\mathbf{q}_{i}^{\top}\right)^{\top}$ and $\phi\left(\mathbf{k}_{i}^{\top}\right)^{\top}$ respectively, the kernel approximation of cell attention is formulated as
\begin{equation}
    \resizebox{0.45\textwidth}{!}{$
    \begin{split}
    \operatorname{Attn}(\mathbf{h}_\text{c})=\widehat{\mathbf{D}}^{-1}\left(\mathbf{Q}^{\prime}\left(\left(\mathbf{K}^{\prime}\right)^{\top} \mathbf{h}_\text{c}\right)\right),   \; \text{with}\; \widehat{\mathbf{D}}=\operatorname{diag}\left(\mathbf{Q}^{\prime}\left(\left(\mathbf{K}^{\prime}\right)^{\top} \mathbf{1}_{N_\text{c}}\right)\right).
    \end{split}
    $}
\end{equation}
With kernel of dimension $r$, the space complexity and time complexity is reduced to $O(N_\text{c} r+N_\text{c} d+r d)$ and $O(N_\text{c} r d)$, respectively. The generalized kernelizable attention boosts our cell transformer to linear space and time complexity, while still delivering results comparable to regular transformers~\cite{choromanski2021rethinking}. The attention mechanism in the model captures the intricate relationships between cells, resulting in an improved representation of the individual cells. 

\subsubsection{Gene Transformer and Protein Transformer}
\label{sec:graph_trans}
 
To leverage biological insights, we include STRING~\cite{szklarczyk2023string} as an addition to provide local information of genes and proteins. Although STRING provides solid prior networks, there may still exist data-related concerns when applied to sequencing data. For instance, the NeurIPS 2022 competition dataset
\footnote{https://www.kaggle.com/competitions/open-problems-multimodal/} 
contains $22,050$ genes, but only $13,101$ of these genes have interactions recorded in the STRING database. This issue also occurs in proteins, as the NeurIPS 2022 competition dataset collects $140$ proteins and only $120$ of them are found within the STRING~\cite{szklarczyk2023string} networks. This means that information about interactions within the remaining molecules is not available unless additional global information is included. To address the concerns, we utilized graph transformers to encode both the local and global information about genes and proteins. Specifically, following GraphGPS~\cite{rampasek2022recipe}, we adapt the graph transformers for gene-gene and protein-protein graphs separately, i.e., gene transformer and protein transformer. 
Since the gene transformer and protein transformer have the same architecture, we will use the gene transformer to illustrate the details. 

A gene transformer layer consists of two parallel components: a message-passing GNN block and a global attention block. The GNN block subtracts the gene interaction information from the prior local network, while the attention block learns global data-specific relationships by allowing each gene to attend to all other genes. The results from two blocks are summed together and then processed by fully connected layers to update the gene embeddings. 
Recall that the adjacency matrix of the gene-gene graph is denoted as $\mathbf{A}_{\text{g}} \in \mathbb{R}^{N_{\text{g}}\times N_{\text{g}}}$, let $\mathbf{h}_g^{\ell} \in \mathbb{R}^{N_{\text{g}} \times d_{\ell}}$ with dimension $d_{\ell}$ be the gene embedding of $\ell$-th layer, the update functions are as follows: 
\begin{equation}
    \resizebox{0.45\textwidth}{!}{$
    \begin{split}
      \mathbf{h}_{\text{g}}^{\ell+1}  =\operatorname{MLP}^{\ell}\left(\mathbf{h}_{{\text{g}},M}^{\ell+1}+\mathbf{h}_{{\text{g}},T}^{\ell+1}\right), \; \text{with} \; \mathbf{h}_{{\text{g}},M}^{\ell+1} =\operatorname{GNN}^{\ell}\left(\mathbf{h}_{\text{g}}^{\ell}, \mathbf{A}_{\text{g}}\right), 
        \mathbf{h}_{{\text{g}},T}^{\ell+1} =\operatorname{Attn}^{\ell}\left(\mathbf{h}_{\text{g}}^{\ell}\right), 
    \end{split}
    $}
\end{equation}
where $\operatorname{GNN}^{\ell}$ and $\operatorname{Attn}^{\ell}$ represent the GNN block and the global attention mechanism; $\operatorname{MLP}^{\ell}$ is a $2$-layer MLP block. The GNN block brings in prior biological insights, while the attention block allows resolving the expressivity bottlenecks caused by over-smoothing~\cite{kreuzer2021rethinking} and over-squashing~\cite{alon2021on}.

Positional encoding (PE) is applied to the gene transformer layer to provide another solution to the expressivity bottlenecks among gene-gene graphs. Message-passing GNNs update gene node embeddings by aggregating local neighborhood representation given by gene knowledge graphs. By incorporating additional positional information, PE helps to differentiate nodes that have the same local surroundings but distinct positions. Accompanied by prior gene-gene interaction networks, PE distinguishes genes by the absolute position of them within the STRING network. This is important from a biological perspective as each gene functions differently. Here, we implement Laplacian PE~\cite{dwivedi2020generalization} and random walk PE~\cite{dwivedi2022graph}. The Laplacian PE captures the spectral information of graph Laplacian by its eigenvectors. Denote the graph Laplacian of input gene graph as $\mathbf{L}_{\text{g}}$, the matrix factorization of $\mathbf{L}_{\text{g}}$ is formulated as
\begin{equation}
    \mathbf{L}_{\text{g}}=\mathbf{I}-\mathbf{D}_{\text{g}}^{-1 / 2} \mathbf{A}_{\text{g}} \mathbf{D}_{\text{g}}^{-1 / 2}=\mathbf{U}_{\text{g}}^{T} \mathbf{\Lambda}_{\text{g}} \mathbf{U}_{\text{g}},
\end{equation}
where $\mathbf{D}_{\text{g}}$ is the degree matrix of gene graph, $\mathbf{\Lambda}_{\text{g}}$ and $\mathbf{U}_{\text{g}}$ correspond to the eigenvalues and eigenvectors respectively. The gene node Laplacian PE of dimension $k$ is defined as the $k$ smallest non-trivial eigenvectors. While Laplacian PE embeds the positional information from the graph Laplacian, the random walk PE tends to grasp the positional information given by the graph clusters. The random walk PE of dimension $k$ is defined with $k$-steps of the random walk
\begin{equation}
    \mathbf{p}_{i}=\left[\mathrm{RW}_{i i}, \mathrm{RW}_{i i}^{2}, \cdots, \mathrm{RW}_{i i}^{k}\right] \in \mathbb{R}^{k},
\end{equation}
where $\mathrm{RW}=\mathbf{A}_{\text{g}} \mathbf{D}_{\text{g}}^{-1}$ is the random walk operator. The term $\mathrm{RW}_{i i}^k$ represents the landing probability of a gene node $i$ to itself after $k$ steps. The processed PE is then combined to gene features through a fully connected layer.

\subsubsection{Cross-Modality Aggregation}
 Transformers are constructed separately for each modality. To build a bridge among those transformers, we implement message passing GNNs among the links which connect nodes of distinct types. The information from cell transformer will denoise the prior knowledge by adding data-specific details into gene embeddings and protein embeddings. Meanwhile, information from gene transformer and protein transformer will bring in biological insights to cell transformer to predict the target proteins. Particularly, we take the advantage of GraphSAGE~\cite{hamilton2017inductive} to transfer the information. Denote the $i$-th destination node as $v_i$ and $j$-th source node as $u_j$.
The information from the source nodes to the destination nodes is updated by:
\begin{equation}
    \resizebox{0.45\textwidth}{!}{$
         \mathbf{h}_{v_i}^{(\ell+1)} = \operatorname{Update}\left(\mathbf{h}^{\ell}_{v_i}, \mathbf{h}^{\ell}_{\mathcal{N}(v_i)} \right) \; \text{with} \; \mathbf{h}_{\mathcal{N}(v_i)}^{\ell} = \operatorname{Aggregate}\left(\left\{\mathbf{h}_{u_j}^{\ell}, \forall\ u_j \in \mathcal{N}(v_i)\right\}\right) 
    $}
\end{equation}
For node $v_i$, the message passing GNN aggregates information from its neighbors through aggregator function ($\operatorname{Aggregate}$). 
The neighborhood information $\mathbf{h}_{\mathcal{N}(v_i)}^{\ell}$ is then combined to the embeddings $\mathbf{h}_{v_i}^{\ell}$ and processed by an updating procedure. The newly generated embeddings are normalized before next iteration. The GNN modules facilitate communication between the transformers, enabling the transformers to leverage various forms of information during the training process.

Now we summarize the workflow of \method{}. As shown in Figure\ref{fig:framework}, we apply transformers within each type of node and utilize message passing GNNs to form the bridges between transformers. In a formal way, for $\ell$-th layer, denote the cell transformer as $\operatorname{Trans}^{\ell}_{\text{c}}$, gene and protein graph transformer as $\operatorname{GT}^{\ell}_{\text{g}}$ and $\operatorname{GT}^{\ell}_{\text{p}}$ respectively. Let $\operatorname{MPG}^{\ell}_{\text{g} \shortrightarrow \text{p}}$ and $\operatorname{MPG}^{\ell}_{\text{p} \shortrightarrow \text{g}}$ be the message passing GNN modules between genes and proteins, and $\operatorname{MPG}^{\ell}_{\text{g} \shortrightarrow \text{c}}$ and $\operatorname{MPG}^{\ell}_{\text{c} \shortrightarrow \text{g}}$ are for the links between genes and cells. We process the cell embeddings with MLP as cell readouts, where $\operatorname{FC}^{\ell}$ represents the $\ell$-th fully connected layer. The updates of node embeddings are achieved by 
\begin{equation}\label{eq:fusion}
    \begin{split}
        \mathbf{h}_{\text{g}}^{\ell+1} =& \operatorname{GT}_{\text{g}}^{\ell}\left(\mathbf{h}_{\text{g}}^{\ell}, \mathbf{A}_{\text{g}}\right)  + \operatorname{MPG}^{\ell}_{\text{p} \shortrightarrow \text{g}}\left(\mathbf{h}_{\text{p}}^{\ell}, \mathbf{S}^T\right) + \operatorname{MPG}^{\ell}_{\text{c} \shortrightarrow \text{g}}\left(\mathbf{h}_{\text{c}}^{\ell}, \mathbf{A}_{\text{RNA}}\right), \\
        \mathbf{h}_{\text{c}}^{\ell+1} =& \operatorname{Trans}^{\ell}_{\text{c}}\left(\mathbf{h}_{\text{c}}^{\ell}\right) + \operatorname{MPG}^{\ell}_{\text{g} \shortrightarrow \text{c}}\left(\mathbf{h}_{\text{g}}^{\ell}, \mathbf{A}_{\text{RNA}}^T\right) + \operatorname{FC}^{\ell}\left(\mathbf{h}_{\text{c}}^{\ell}\right), \\
        \mathbf{h}_{\text{p}}^{\ell+1} =& \operatorname{GT}_{\text{p}}^{\ell}\left(\mathbf{h}_{\text{p}}^{\ell}, \mathbf{A}_{\text{p}}\right) + \operatorname{MPG}^{\ell}_{\text{g} \shortrightarrow \text{p}}\left(\mathbf{h}_{\text{g}}^{\ell}, \mathbf{S}\right), 
    \end{split}
\end{equation}
where $\mathbf{S}$, $\mathbf{A}_{\text{p}}$, $\mathbf{A}_{\text{g}}$, $\mathbf{A}_{\text{RNA}}$ are adjacency blocks defined in Section~\ref{sec:graph_const}. 

\textbf{Prediction Layer.}
With the number of layers be $L$, the predictions are given by one extra full connected layer $\operatorname{FC}^{L +1}$ as
\begin{equation}
    \begin{split}
      \widehat{\mathbf{X}}_{\text{p}} = \operatorname{FC}^{L + 1}\left(\mathbf{h}_{\text{c}}^{L+1} \right), \; \text{with} \;  \mathbf{h}_{\text{c}}^{L+1} = \operatorname{Concat}\left(\mathbf{h}_{\text{c}}^{\ell},\ \ell\ \text{in}\ \{1, 2, \dots, L\} \right),
    \end{split}
\end{equation}
where $\widehat{\mathbf{X}}_{\text{p}}\in \mathbb{R}^{N_{\text{c}}\times N_{\text{p}}}$ is the final prediction. To optimize the framework, we adapt a Mean Square Error (MSE) loss to measure the difference between the predictions and the ground-truth values:
\begin{equation}
    \mathcal{L} \left(\widehat{\mathbf{X}}_{\text{p}}, \mathbf{X}_{\text{p}}\right) = \frac{1}{N_p N_c} \sum_{i=1}^{N_p}\sum_{i=1}^{N_c}\left(\mathbf{X}_{\text{p}}^{ij}-\widehat{\mathbf{X}}_{\text{p}}^{ij}\right)^{2} .
\end{equation}

%% file: sections/experiment.tex
\section{Experiment}
In this section, we present the experimental results of \method{} against baselines on benchmark datasets. In particular, we aim to answer the following questions:
\begin{itemize}[leftmargin=*]
    \item \textbf{RQ1:} How does \method{} perform compare against baselines based on various evaluation metrics?
    \item \textbf{RQ2:} Given various choices of the PEs, how do they affect the performance of \method{}?
    \item \textbf{RQ3:} What is the best way to fuse information across modalities?
    \item \textbf{RQ4:} Can \method{} handle other modality prediction tasks?
    \item \textbf{RQ5:} Can \method{} process large datasets?
    \item \textbf{RQ6:} How does each of the model component impact the performance of \method{}?
\end{itemize}
Before presenting our experimental results and observations, we first introduce the experimental settings.

\subsection{Experimental Settings}
\subsubsection{Datasets}
We follow the setting of the NeurIPS multimodal single-cell integration competition of the year 2021~\cite{luecken2021sandbox} and 2022
and collect the joint measurements of gene expression and surface protein levels datasets from the competitions. Both datasets contain the raw counts, which represent the number of reads per gene per cell, as well as the normalized counts. For the NeurIPS 2021 competition, we pick the data corresponding to the task of protein abundance prediction via gene expression and refer to it as ``GEX2ADT''. The processed data is centered and log-transformed for denoising purposes. For the competition in 2022, which we refer to as ``CITE'', the objective is to utilize CITE-seq~\cite{stoeckius2017simultaneous} data measured from days 2, 3, and 4 to predict the protein level on day 7 from different individuals. It is worth mentioning that the protein level testing data is not available during the completion of this work. Therefore, we simulate the competition scenario by treating the training data from day 4 as our testing set. The processed RNA data is centered and log-transformed, while the normalized protein levels are denoised and scaled by background~\cite{kotliarov2020broad}. We summarize the dataset details in Table~\ref{table:dataset} in Appendix~\ref{app:exp}. 

\subsubsection{Baselines}
We evaluate the performance of \method{} against state-of-art multimodal prediction models among the task of using gene expression to predict surface protein levels. The selected baselines are as follows:
\begin{itemize}[leftmargin=*]
    \item \textbf{Cross-modal Autoencoders}~\cite{yang2021multi}, short for CMAE, incorporated multiple autoencoders to integrate multimodal data and utilized domain knowledge by adding discriminative loss to the training process to align shared markers or clusters.
    \item \textbf{BABEL}~\cite{wu2021babel} proposed a general framework for multimodal translation with modality-specific encoders and decoders. Note that initially, BABEL focused on RNA and ATAC-seq~\cite{buenrostro2013transposition} data. In this evaluation, we repurpose BABEL to the RNA to protein setting.
    \item \textbf{scMM}~\cite{minoura2021mixture} modeled the multimodal data with generative setting. We note that the input of scMM is restricted to raw counts, and the predictions are scaled as centered log-transformed data.
    \item \textbf{ScMoGNN}~\cite{wen2022graph} involved domain knowledge like biological pathways to enhance the GNNs. The original ScMoGNN followed a transductive setting. In this work, we implement an inductive setting of ScMoGNN for a fair comparison with the baselines.
\end{itemize}

\subsubsection{Parameter Setting}
To benchmark the performance of baselines and \method{}, we uniformly employ inductive settings among both datasets. On the CITE, we use the data measured on day 4 for testing and randomly split $80/20\%$ of the data prior to day 4 for training and validation. On the GEX2ADT, we randomly pick $15\%$ of the training data for validation and evaluate the predictions on the testing set. For BABEL, the hidden dimension is tuned from $\{16, 32, 64, 128\}$. For CMAE, the weights of adversarial loss and reconstruction loss are chosen from $\{0.1, 1, 2.5, 5, 10\}$. For scMM, the hidden dimensions are tuned from $\{16, 32, 64, 128\}$. For ScMoGNN, the weight decay parameter of the optimizer is tuned from $\{5\times 10^{-6},1 \times 10^{-5}, 5 \times 10^{-5}, 1 \times 10^{-4}\}$.

\subsection{Evaluation of Predictions}
\input{tables/main_results.tex}
We evaluate the final protein-level prediction performance using Root Mean Square Error (RMSE) and Mean Absolute Error (MAE). Meanwhile, because multimodal data usually suffers from the influence of batch effects and unbalanced measuring depth, the count's scale of each cell may vary significantly, which will substantially affect the RMSE and MAE metrics. Therefore, we also include the Pearson correlation coefficient (Corr), which is cell-wise normalized by the mean and variance of the input, as a robust and scale-free metric to evaluate the predictions. A lower RMSE or MAE score indicates a geometrically closer estimation of the protein levels, while a higher Corr score suggests a statistically more similar match to the actual value. We report the mean and the standard deviation of each metric across five different runs, and the results are illustrated in Table~\ref{table:main}. The best performance is highlighted in bold. 

To answer the first question, we note that our \method{} consistently outperforms all other baselines according to all three metrics on both datasets, indicating that \method{} successfully captures the quantitative characteristics of target protein levels given the input gene expression measurements. Particularly, for the CITE, \method{} achieved significantly lower RMSE compared to the second-best model ScMoGNN, by $0.04$. More importantly, \method{} achieved a significant improvement in terms of the Pearson correlation metrics over all other baselines, with a noticeably lower performance variation across runs.

\begin{figure}[htb]
    \centering
    \vspace{-1.2em}
    \caption{Least well-predicted protein comparison.}
    \includegraphics[width=0.47\textwidth]{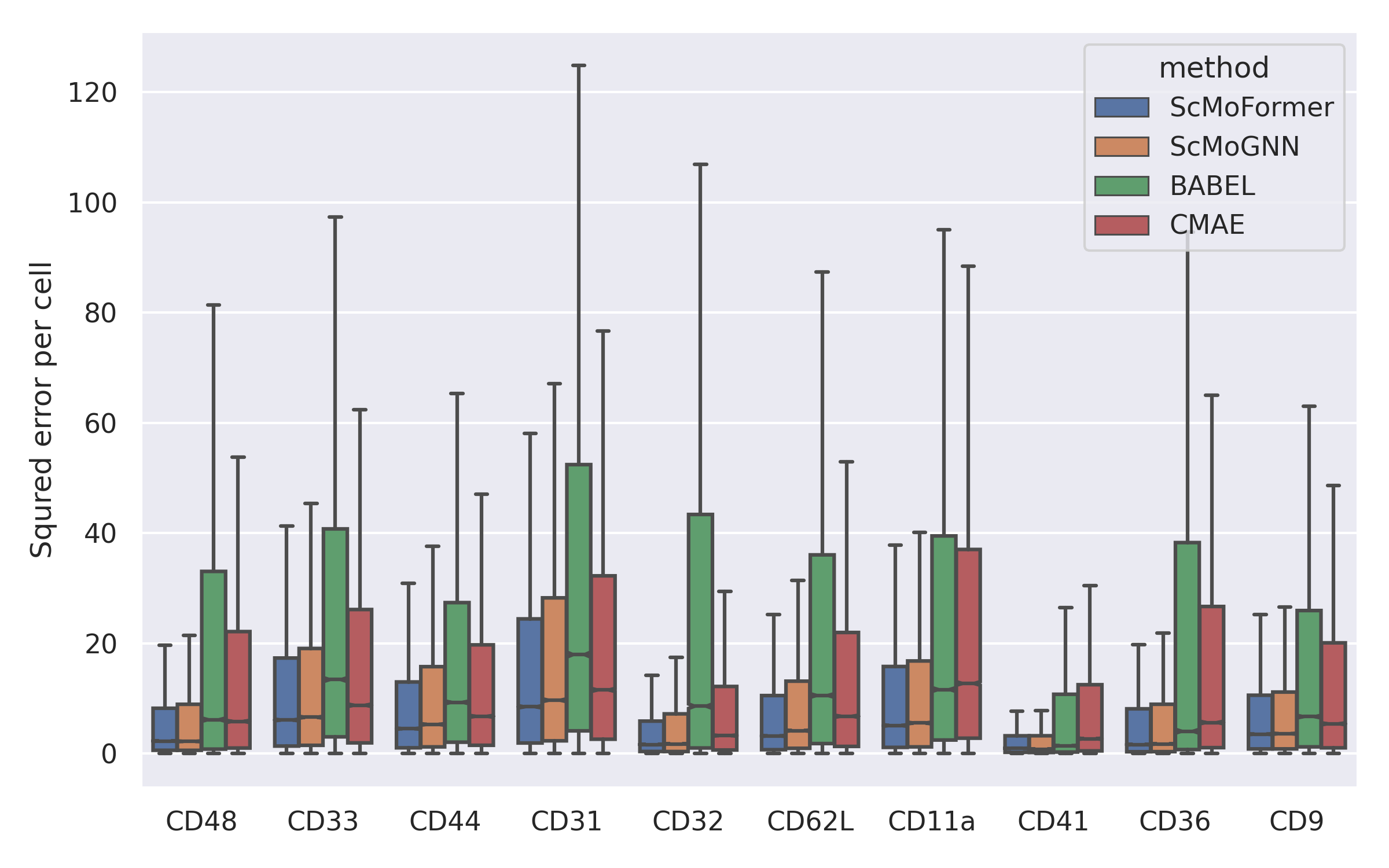}
    \vspace{-1.2em}
    \label{fig:bot10comp}
\end{figure}

We further analyze the performance of different models on proteins that are least well captured by any models. Specifically, for each model, we compute the RMSE for each protein separately and identify ten proteins that resulted in the highest average RMSE across all models. As shown in Figure~\ref{fig:bot10comp}, \method{} and ScMoGNN achieved relatively stable results and are consistently better compared to BABEl and CMAE.

\vspace{-0.5em}
\subsection{Positional Encoding}
\label{sec:pe}
As mentioned in Section~\ref{sec:graph_trans}, we implement Laplacian PE~\cite{dwivedi2020generalization} and random walk PE~\cite{dwivedi2022graph} to capture positional information of prior knowledge graph. For ease of notation, we use the abbreviation PE to refer to the PE. To benchmark the impact of the two types of PE among two datasets and answer the second question, we show the performance of \method{} with each PE and compare them with the scenario without any PE. The mean and standard deviation of RMSE scores of five runs are shown in Table~\ref{table:pe}. 

According to the results, the influence of PE varies among datasets. \method{} reaches the best RMSE score on CITE without PE, while two types of PE both improve the performance on GEX2ADT. Notice that in Table~\ref{table:dataset}, the RNA zero rate of CITE is significantly lower compared to the GEX2ADT, providing the model with greater access to data-specific information. If the data contains sufficient information, then the neighborhood information from the GNNs alone is adequate and there is no need for the extra prior knowledge from the PEs. This is further supported by the observation that random walk PE performs better than Laplacian PE in both datasets. The Laplacian PE models global information by using the spectral information of the graph Laplacian, while the random walk PE encodes local information by accessing the landing probability of a $k$-step random walk. In cases where the prior knowledge may be noisy for downstream tasks, the local information alone is enough for predictions and the global structure becomes redundant.

\input{tables/pe.tex}

\subsection{Comparasion of Different Fusion Strategies}

Our framework is based on hybrid fusion. Layer-wise speaking, we refer to our fusion strategy as \textit{concurrent} fusion since node representations simultaneously go through GNNs and transformers. In the case of protein nodes, information from protein nodes and gene nodes is fused after being processed by protein transformers and gene-protein (gene to protein) GNNs, respectively. In the following experiment, we compared the performance of layer-wise \textit{concurrent} fusion, \textit{GNN-first} fusion, and \textit{mixed} fusion. Still, in the case of protein nodes, we implemented \textit{GNN-first} fusion by first summing protein embeddings and the gene-protein GNNs outputs, and then passing through protein transformers. For \textit{mixed} fusion, we utilized \textit{concurrent} fusion on protein and gene nodes while applying \textit{GNN-first} fusion on cell nodes. We summarize the prediction RMSEs and the standard deviations across five runs in Table~\ref{tab:fusion}.

\begin{table}[h]
\centering
\vspace{-0.5em}
\caption{The prediction RMSEs for different Fusion strategies.}\label{tab:fusion}
\vspace{-1.2em}
\resizebox{0.48\textwidth}{!}{
\begin{tabular}{@{}lccc@{}}
\toprule
        & Concurrent          & GNN-first           & Mixed               \\ \midrule
CITE    & \textbf{1.62720}$\pm$\textbf{0.00731} & 1.63054$\pm$0.01018 & 1.63092$\pm$0.00631 \\
GEX2ADT & \textbf{0.41987}$\pm$\textbf{0.00234} & 0.42861$\pm$0.00094 & 0.42948$\pm$0.00158 \\ \bottomrule
\end{tabular}
}
\vspace{-1.2em}
\end{table}

As presented in the table, the original \textit{concurrent} fusion strategy exhibited superior performance compared to the other two fusion strategies. The \textit{concurrent} setting allows each modality to utilize intra-modal information prior to combining with other modalities, resulting in better performance since each modality holds varying importance for downstream tasks. However, the improvement in performance was not significant enough for the CITE. This observation is consistent with the findings of other experiments. Since the RNA zero rate in the CITE is considerably lower than that of the GEX2ADT, the significance of data-specific information (cell nodes) in the CITE outweighs the importance of other modalities, resulting in a relatively smaller performance boost.

\subsection{Handling Other Single-Cell Multimodal Prediction Tasks}

In this work, we focused on the specific task GEX2ADT (gene expression to protein levels) as a showcase. However, our framework is versatile and can be applied to other modality prediction tasks, i.e., the other three tasks mentioned in the NeurIPS 2021 competition~\cite{luecken2021sandbox} including ADT2GEX (protein levels to gene expression), GEX2ATAC (gene expression to chromatin accessibility) and ATAC2GEX (chromatin accessibility to gene expression). 

\noindent \textbf{ADT2GEX:} In our framework, we constructed a multimodal heterogeneous graph consisting of gene, protein and cell nodes. For the GEX2ADT task, we removed the cell-protein edges to eliminate information leakage. Similarly, for the ADT2GEX task, we incorporate protein measurements into the cell-protein edges while removing the cell-gene edges. Cell embeddings were initialized by the reduced protein levels, and protein embeddings were initialized by the weighted sum of cell embeddings, where the weights are with the normalized protein levels. 

\noindent \textbf{GEX2ATAC and ATAC2GEX:} In the context of the GEX2ATAC and ATAC2GEX tasks, we removed the protein nodes from the multimodal graph. To initialize the cell embeddings for the GEX2ATAC task, we used the reduced gene expression values, while for gene embeddings, we computed a weighted sum of cell embeddings using normalized gene measurements as weights. For the ATAC2GEX task, we masked the cell-gene edges in the testing set. We initialized the cell embeddings using the reduced ATAC measurements, and the gene embeddings were randomly initialized.

We have compared the performance of \method{} with scMoGNN on the four tasks in Table~\ref{tab:others}. Note that \method{} outperformed scMoGNN in tasks that involve protein modality and vice versa. These results suggest that incorporating prior information of protein nodes can enhance the performance of \method{} when protein modality is present. The RMSE scores across five runs are summarized as in Table~\ref{tab:others}.

\begin{table}[h]
\centering
\vspace{-0.5em}
\caption{Results on other Single-Cell Mutimodal Tasks.}\label{tab:others}
\vspace{-1.2em}
\resizebox{0.48\textwidth}{!}{
\begin{tabular}{l|cccc}
\toprule
           & GEX2ADT             & ADT2GEX             & GEX2ATAC            & ATAC2GEX            \\ \midrule
scMoFormer & \textbf{0.4198}7$\pm$\textbf{0.00234} & \textbf{0.31547}$\pm$\textbf{0.00184} & 0.17885$\pm$0.00008 & 0.23988$\pm$0.00031 \\
scMoGNN    & 0.42576$\pm$0.01180 & 0.32250$\pm$0.00136 & \textbf{0.17823}$\pm$\textbf{0.00011} & \textbf{0.23021}$\pm$\textbf{0.00219} \\ \bottomrule
\end{tabular}
}
\vspace{-1.8em}
\end{table}

\subsection{Training Efficiency of \method{}} \label{sec:efficiency}
Since scMoGNN is the best-performing baseline, we compared the running time and total GPU memory of \method{} and scMoGNN across five runs on one Quadro RTX 8000 GPU. The results are summarized in Table~\ref{tab:efficiency}. We observed that with higher GPU consumption, \method{} required a significantly shorter running time compared to scMoGNN. It is worth noting that \method{} was configured with a relatively large batch size of 8000 cells per batch and a hidden dimension of 512, in order to achieve better performance and better utilization of computing resources. One can surely reduce the required GPU memory of \method{} by limiting the setting accordingly. 

\begin{table}[h]
\centering
\vspace{-0.5em}
\caption{Efficiency Comparison}\label{tab:efficiency}
\vspace{-1.2em}
\resizebox{0.48\textwidth}{!}{
\begin{tabular}{@{}l|cc|cc@{}}
\toprule
\multicolumn{1}{l|}{} & \multicolumn{2}{l|}{\textbf{Running time (min)}} & \multicolumn{2}{l}{\textbf{GPU memory (GB)}}         \\ \midrule
                      & CITE                   & GEX2ADT                 & CITE                   & \multicolumn{1}{l}{GEX2ADT} \\
\method{}            & 24.82$\pm$4.16         & 17.95$\pm$3.56          & \multicolumn{1}{r}{38} & 21                          \\
scMoGNN               & 58.89$\pm$7.77         & 108.54$\pm$21.22        & \multicolumn{1}{r}{26} & 12                          \\ \bottomrule
\end{tabular}
}
\vspace{-0.8em}
\end{table}

\noindent \textbf{Can \method{} process large datasets?} In practical applications, it exhibits the capability to process large datasets. We address this issue from the perspectives of both the model and the data:
\begin{itemize}[leftmargin=*]
\item \textbf{Data aspect:} Due to technological or biological reasons, the number of RNA nodes and the number of protein nodes will be similar across datasets. Therefore, large datasets mean more cells, which can be handled by mini-batching cells.
\item \textbf{Model aspect:} In our multimodal transformer module, we employed linearized transformers~\cite{choromanski2021rethinking} with linear space and time complexity, which can be conveniently adapted to large datasets. Concerning GNNs, we incorporated GraphSAGE~\cite{hamilton2017inductive}, whose space and time complexity are related to the number of edges and hidden layer dimensions. Indeed, we can control the number of edges by mini-batching cells and make appropriate adjustments based on available computational resources.
\end{itemize}

\subsection{Ablation Study}
Table~\ref{table:main} demonstrates that models that incorporate domain knowledge perform better in modality prediction compared to those that do not. BABEL, ScMoGNN, and \method{} are the three models that make use of domain knowledge, and they show improved performance compared to the other two models. Among these three models, ScMoGNN, which is based on GNNs, performs better than BABEL, while \method{} outperforms all other models with its combination of transformers and GNNs framework. Given that \method{} includes a multimodal heterogeneous graph and three transformers, this raises the questions: \textit{Why no cell-cell graph and cell-protein graph? Which transformer has the biggest impact on performance? How much do the transformers contribute to the improvement in performance?}

\subsubsection{Exclusion of Cell-Cell Graph and Cell-Protein Graph.}\label{app:remark}
In Remark of Section~\ref{sec:subgraph}, we provided an explanation for our decision to exclude the cell-cell and cell-protein graphs. Additionally, through our empirical analysis, we observed a decrease in performance on both datasets when these links were incorporated into our heterogeneous graph. In comparison to the original w/o neither graph setting, the performance drop was evident in both the w/ cell-protein graph and w/ cell-cell graph settings, with the former demonstrating a more significant decline. The prediction RMSEs and standard deviations of five runs are summarized in Table~\ref{tab:graph_variants}.

\begin{table}[h]
\vspace{-0.5em}
\caption{The prediction RMSEs on different graph variants.}\label{tab:graph_variants}
\vspace{-1.2em}
\resizebox{0.48\textwidth}{!}{
\begin{tabular}{@{}lcccc@{}}
\toprule
        & w/o neither graph   & w/ cell-cell graph  & w/ cell-protein graph & w/ both graphs      \\ \midrule
CITE    & \textbf{1.62720}$\pm$\textbf{0.00731} & 1.68932$\pm$0.00751 & 1.71742$\pm$0.01692   & 1.70094$\pm$0.02207 \\
GEX2ADT & \textbf{0.41987}$\pm$\textbf{0.00234} & 0.42441$\pm$0.00094 & 0.43911$\pm$0.00414   & 0.42983$\pm$0.00231 \\ \bottomrule
\end{tabular}
}
\vspace{-1.2em}
\end{table}

\noindent \textbf{Why cell-cell graph did not help:} In the above experiment, we constructed a k-NN cell-cell graph by measuring the similarity in gene expression between cells. Nevertheless, since gene expression measurements often suffer from noise and sparsity, this static cell-cell graph may have introduced biased information into the downstream task. To address this issue, we utilized a cell transformer module to learn the dynamic cell-cell interactions via multi-head attention mechanism, which led to improved performance.

\noindent \textbf{Why cell-protein graph did not help:} For the cell-protein links, we incorporated target surface protein levels into edge weights, similar to how we built the cell-gene graph. However, these links conveys information about the prediction targets of the training set, causing the model to overfit easily. Hence, eliminating the cell-protein links served to eradicate information leakage.

\subsubsection{Influence of Every Transformer} 
\begin{figure}[htb]%
    \vspace{-2.em}
    \centering
    \caption{RMSE$\downarrow$ results of keeping only one Transformer.}%
    \vspace{-1.2em}
    \subfloat[CITE]{{\includegraphics[width=0.5\linewidth]{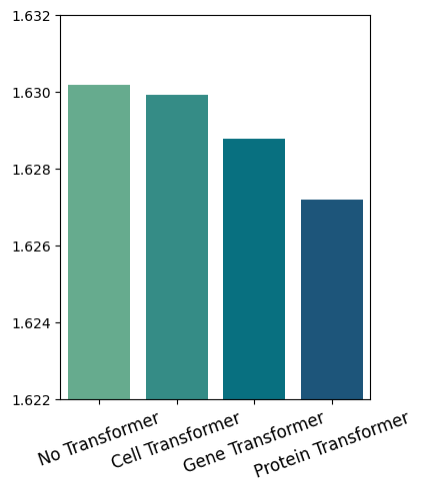} }}%
    \subfloat[GEX2ADT]{{\includegraphics[width=0.5\linewidth]{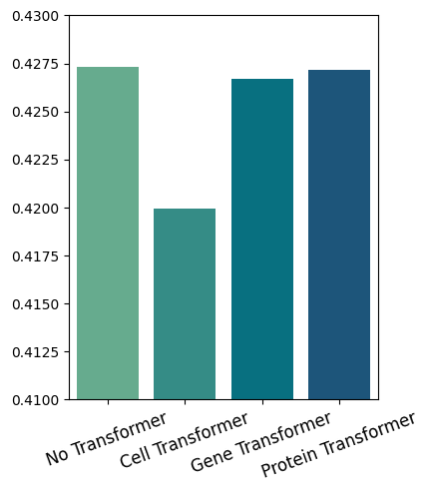} }}%
    \label{fig:one_trans}%
    \vspace{-1.8em}
\end{figure}

The propose multimodal transformers consist of three different transformers, namely the cell transformer, gene transformer and protein transformers. As our predictions are based on the cell readout, it is expected that each of the three transformers will have different levels of impact on the performance. To quantify the specific impact of a single transformer, we conduct an experiment by removing the other two transformers and measuring the prediction RMSE scores. The results of the evaluation, including the scores of three partial models and the model with no transformers, are summarized in 
Figure~\ref{fig:one_trans}.

The performance of the gene transformer and protein transformer is better than that of the cell transformer in the CITE, while it is the opposite in the GEX2ADT. This can be explained by the difference in RNA zero rate between the two datasets, as shown in Table~\ref{table:dataset}. For the GEX2ADT, the high RNA zero rate means less information, making the cell transformer crucial in increasing performance by drawing more information from the data. On the other hand, the CITE has a lower zero rate, meaning it provides more information, allowing the gene transformer and protein transformer to enhance the model by adding external biological knowledge.

\subsubsection{How to Utilize Prior Knowledge} To answer the third question, we compare \method{} with two GNN-based models in Table~\ref{table:ablation}. The model "GNN-prior" refers to the GNNs that built on the same graph in Section~\ref{sec:graph_const}, while the "GNN" model is constructed using only the cell-gene graph without incorporating any prior information. The results show that the incorporation of prior knowledge into the graph results in a slight performance boost in both datasets. However, when multimodal transformers are included, the performance improvement is much more pronounced. This highlights the usefulness of prior knowledge and the importance of transformers to effectively incorporate this information into the model.
\input{tables/ablation.tex}

%% file: tables/main_results.tex
\begin{table*}[t] 
\small
\caption{Prediction evaluations based on different metrics (score $\pm$ std).}
\vspace{-1em}
\label{table:main}
\begin{threeparttable}
\begin{tabular}{c|ccc|ccc}
    \toprule
    Dataset      & \multicolumn{3}{|c|}{CITE}  & \multicolumn{3}{|c}{GEX2ADT}\\
    Metric       & RMSE $\downarrow$                 & MAE $\downarrow$                   & Corr   $\uparrow$               & RMSE $\downarrow$                 & MAE $\downarrow$                  & Corr  $\uparrow$                \\ 
    \midrule 
    BABEL        & 1.67388 $\pm$ 0.00765 & 1.07777 $\pm$ 0.00602 & 0.87475 $\pm$ 0.00119 & 0.45387 $\pm$ 0.00738 & 0.30720 $\pm$ 0.00618 & 0.86144 $\pm$ 0.00274 \\
    CMAE         & 2.00874 $\pm$ 0.02088 & 1.21897 $\pm$ 0.01042 & 0.82502 $\pm$ 0.00843 & 0.51549 $\pm$ 0.00857 & 0.34855 $\pm$ 0.00488 & 0.81565 $\pm$ 0.00463\\
    scMM\footnote{}         & -                     & -                     & -                     & 0.64067 $\pm$ 0.00722 & 0.43407 $\pm$ 0.00307 & 0.68287 $\pm$ 0.00981\\
    ScMoGNN      & 1.66634 $\pm$ 0.00741 & 1.07577 $\pm$ 0.00372 & 0.87788 $\pm$ 0.00113 & 0.42576 $\pm$ 0.01180 & 0.28819 $\pm$ 0.00976 & 0.87051 $\pm$ 0.00524\\
    scMoFormer & \textbf{1.62720 $\pm$ 0.00731} & \textbf{1.05639 $\pm$ 0.00221} & \textbf{0.88552 $\pm$ 0.00080} & \textbf{0.41987 $\pm$ 0.00234} & \textbf{0.28289 $\pm$ 0.00223} & \textbf{0.87698 $\pm$ 0.00121}\\
    \bottomrule
\end{tabular}
\begin{tablenotes}
   \item[5] The scale of scMM predictions is not compatible with that of normalized protein levels.
  \end{tablenotes}
\end{threeparttable}
\vspace{-1em}
\end{table*}

%% file: tables/pe.tex
\begin{table}[tb]
    \centering
    \vspace{-1em}
    \caption{Prediction RMSE results of different positional encoding (score $\pm$ std).}
    \vspace{-1.2em}
    \scalebox{1.}{
    \begin{tabular}{c|c|c}
        \toprule
        & \textbf{CITE} & \textbf{GEX2ADT} \\ \midrule
        Laplacian PE & 1.63161 $\pm$ 0.01082 &  0.42025 $\pm$ 0.00243 \\
        Random Walk PE & 1.63014 $\pm$ 0.01129 & \textbf{0.41987 $\pm$ 0.00234} \\
        w/o PE & \textbf{1.62720 $\pm$ 0.00731} & 0.42202 $\pm$ 0.00399 \\ 
        \bottomrule
    \end{tabular}
    }
    \vspace{-1.5em}
    \label{table:pe}
\end{table}

%% file: tables/ablation.tex
\begin{table}[tb]
    \centering
    \caption{Prediction RMSE results of scMoFormer over GNNs (score $\pm$ std).}
    \vspace{-1.2em}
    \scalebox{1.}{
    \begin{tabular}{c|c|c}
        \toprule
        & \textbf{CITE} & \textbf{GEX2ADT} \\ \midrule
        GNN & 1.63071 $\pm$ 0.01081 & 0.43070 $\pm$ 0.00237 \\
        GNN-prior & 1.63020 $\pm$ 0.00672 & 0.42731 $\pm$ 0.00138 \\
        scMoFormer & \textbf{1.62720 $\pm$ 0.00731} & \textbf{0.41987 $\pm$ 0.00234} \\  
        \bottomrule
    \end{tabular}
    }
    \vspace{-1.8em}
    \label{table:ablation}
\end{table}

%% file: sections/conclusion.tex
\section{Conclusion}
Recent advances in multimodal single-cell technology have enabled the simultaneous profiling of the transcriptome alongside other cellular modalities, leading to an increase in the availability of multimodal single-cell data. In this paper, we present \method{}, a multimodal transformer model for single-cell surface protein abundance from gene expression measurements. We combined the data with prior biological interaction knowledge from the STRING database into a richly connected heterogeneous graph and leveraged the transformer architectures to learn an accurate mapping between gene expression and surface protein abundance. Remarkably, \method{} achieves superior and more stable performance than other baselines on both 2021 and 2022 NeurIPS single-cell datasets.

\noindent\textbf{Future Work.}
Our framework of multimodal transformers with the cross-modality heterogeneous graph goes far beyond the specific downstream task of modality prediction, and there are lots of potentials to be further explored. Our graph contains three types of nodes. While the cell embeddings are used for predictions, the remaining protein embeddings and gene embeddings may be further interpreted for other tasks. The similarities between proteins may show data-specific protein-protein relationships, while the attention matrix of the gene transformer may help to identify marker genes of each cell type. Additionally, we may achieve gene interaction prediction using the attention mechanism.
To extend more on transformers, a potential next step is implementing cross-attention cross-modalities. Ideally, all three types of nodes, namely genes, proteins, and cells, would be jointly modeled using a large transformer that includes specific regulations for each modality.

%% file: sections/acknowledgments.tex
\section{Acknowledgement}
This research is supported by the National Institutes of Health (NIH) under grant number UO1 DE029255 and R01 DE026728, the National Science Foundation (NSF) under grant numbers CNS 2246050, IIS1845081, IIS2212032, IIS2212144, IOS2107215, DUE 2234015, DRL 2025244 and IOS2035472, the Army Research Office (ARO) under grant number W911NF-21-1-0198, the Home Depot, Cisco Systems Inc, Amazon Faculty Award, Johnson \& Johnson, JP Morgan Faculty Award and SNAP.

%% file: sections/appendix.tex


\section{Experimental Details} 
\label{app:exp}
We summarize the dataset details in Table~\ref{table:dataset}. The experimental results on CBMC are shown in Appendix~\ref{app:cbmc}.
\begin{table}[htb]
    \centering
    \vspace{-1.em}
    \caption{Dataset Statistics.}
    \vspace{-1.2em}
    \scalebox{1.}{
    \begin{tabular}{c|ccc}
        \toprule
        & \textbf{CITE} & \textbf{GEX2ADT} & \textbf{CBMC}\\ \midrule
        Number of RNA & 22,050 & 13,953 & 20,501 \\ 
        Number of Proteins & 140 & 134 & 10\\
        Train Cells & 42,843 & 66,175 & 7755 \\
        Test Cells & 28,145 & 1,000 & 916 \\
        RNA Zero Rate & 0.780 & 0.904 & 0.953 \\
        \bottomrule
    \end{tabular}
    }
    \vspace{-1.8em}
    \label{table:dataset}
\end{table}


\section{More Experimental Results}

\noindent \textbf{Performance on CBMC CITE-seq Dataset.}
\label{app:cbmc}
We have conducted additional experiments on the CBMC CITE-seq dataset obtained from Seurat~\cite{hao2021integrated}, which contains 20501 RNAs and 10 proteins. For the purpose of comparison, we employed scMoGNN along with \method{} since the former is the best performing baseline. The results of five runs are presented in Table~\ref{tab:cbmc}. Our findings show that \method{} consistently outperformed scMoGNN across all three metrics, which aligns with our previous observations on the CITE and GEX2ADT datasets.

\begin{table}[h]
\centering
\vspace{-0.5em}
\caption{Results on CBMC dataset.}\label{tab:cbmc}
\vspace{-1.2em}
\resizebox{0.48\textwidth}{!}{
\begin{tabular}{@{}lccc@{}}
\toprule
CBMC       & \textbf{RMSE}                & \textbf{MAE}                 & \textbf{Corr}                \\ \midrule
scMoGNN    & 0.60699$\pm$0.01400 & 0.43439$\pm$0.00708 & 0.95043$\pm$0.00334 \\
\method{} & \textbf{0.57272}$\pm$\textbf{0.02660} & \textbf{0.41790}$\pm$\textbf{0.01831} & \textbf{0.9541}8$\pm$\textbf{0.00381} \\\bottomrule
\end{tabular}
}
\vspace{-1.8em}
\end{table}